\documentstyle[12pt]{article}
\begin{document}
\begin{center}
\section*{\bf The phase coherence of light from extragalactic sources - 
direct evidence
against first order Planck scale fluctuations in time and space}
\vspace{1.5mm}

Richard Lieu and Lloyd W. Hillman
 
Department of Physics, University of Alabama, Huntsville,
AL 35899, U.S.A.
\end{center}
 
\vspace{1.5mm}
 
\noindent
{\bf Abstract}

We present a method of directly testing whether time  continues to
have its usual meaning 
on scales of $\leq t_P = (\hbar G/c^5)^{\frac{1}{2}} \approx
5.4 \times 10^{-44}$ s, the Planck time.
According to quantum gravity, the time $t$ of an event
cannot be determined more accurately than a standard deviation of
the form $\sigma_t/t = a_o (t_P/t)^\alpha$, where $a_o$ and $\alpha$
are positive constants $\sim 1$; likewise distances  are subject to
an ultimate
uncertainty $c \sigma_t$, where $c$ is the speed of light.
As a consequence, the period and wavelength of light cannot be specified
precisely; rather, they are
independently subject
to the same intrinsic limitations in our knowledge of time and space, so
that even the most monochromatic plane wave must in reality be
a superposition of waves with varying $\omega$ and ${\bf k}$, each having a
different phase velcocity $\omega/k$.  For the entire accessible range of the
electromagnetic spectrum this effect is extremely small, but can
cumulatively lead to a complete loss of phase information if the emitted
radiation propagated a sufficiently large distance.
Since, at optical frequencies, the phase coherence of light from a distant
point source is a necessary condition for the presence of diffraction 
patterns when the source is viewed through a telescope, such observations
offer by far the most sensitive and uncontroversial test.   We show that
the HST detection of Airy rings from the active galaxy PKS1413+135, located
at a distance of 1.2 Gpc, secures the exclusion of all 
first order ($\alpha = 1$)
quantum gravity fluctuations with an amplitude $a_o > 0.003$.  The same result
may be used to deduce that the speed of light in vacuo is exact
to a few parts in $10^{32}$.

\vspace{3mm}

\noindent
1. INTRODUCTION - THE PLANCK SCALE

It is widely believed that time ceases to be well-defined
at intervals $\leq t_P$,
where quantum fluctuations
in the vacuum metric tensor renders General Relativity an
inadequate theory.  Both $t_P$ and its corresponding distance
scale $l_P = c t_P$, the Planck length, play a vital role in
the majority of theoretical models (including superstrings) 
that constitute innumerable papers
attempting to explain how the universe was born, and how it evolved during
infancy (see e.g, Silk 2000 and references therein).  Given this
background, we desperately lack
experimental data that reveal even the slightest anomaly in
the behavior of time and space at such
small scales.
Although the recent efforts in utilizing
gravitational wave interferometry and the observation of
ultra-high energy (UHE) quanta carry potential
(Amelino-Camelia 2001, Ng et al 2001,
Lieu 2002),
they are still some way from delivering a verdict, because the
conclusions are invariably subject to interpretational issues.
Here we wish to describe how an entirely different
yet well established technique has hitherto been overlooked: not only would it
enable direct tests for Planck scale fluctuations
(and revealing the detailed properties of any such effects),
but also the measurements performed to date could already be used to
eliminate prominent theories.

Owing to the variety of proposed models we begin by describing the
common feature
that define the phenomenon being searched:
if  a time $t$ is so small that $t \rightarrow t_P$
even the best clock ever made will only be able to determine it
with an uncertainty $\delta t \geq t$.  To express this
mathematically we may write the intrinsic
standard deviation of time as $\sigma_t/t = f(t_P/t)$, where $f \ll 1$ for
$t \gg t_P$ and $f \geq 1$ for $t \leq t_P$.  Over the range $t \gg t_P$
the (hitherto unknown) function $f$ can be expanded as follows:
\begin{equation}
f(x) = x^\alpha (a_o + a_1 x + a_2 x^2 + ...) \approx a_o x^\alpha~
{\rm for}~x \ll 1,
\end{equation}
where $x = t_P/t$,  and both $a_o$,
$\alpha$  are positive constants ($\sim 1$ for all reasonable scenarios).
Since for the rest of this paper we shall be concerned only with
times $t \gg t_P$  we may take an approximate form of Eq. (1) as:
\begin{equation}
\frac{\sigma_t}{t} \approx a_o \left(\frac{t_P}{t}\right)^\alpha
\end{equation}

Our appreciation of how
Eq. (2) may affect measurements
of angular frequencies and wavevectors $(\omega, {\bf k})$  arises
from the realization that if a quantity
$\omega > \omega_P = 2 \pi /t_P$
can be determined accurately such a calibration will lead to a
`superclock' that keeps time to within $\delta t < t_P$.  Thus
as $\omega \rightarrow \omega_P$, $\omega$ should fluctuate randomly such that
$\delta \omega/\omega \rightarrow 1$.  Indeed
for the case of $\sigma_t \approx t_P$ (i.e. Eq. (2) 
with $\alpha =1$) the following Eq. was shown by Lieu (2002)
to be an immediate consequence:
\begin{equation}
\frac{\sigma_{\omega}}{\omega}
\approx a_o \frac{\omega}{\omega_P},~{\rm or}~~
\frac{\sigma_{E}}{E} \approx a_o \frac{E}{E_P},
\end{equation}
where $E = \hbar \omega$ and $E_P = \hbar \omega_P =
h/t_P \approx 8.1 \times 10^{28}$ eV.
Further, for any value of $\alpha$ it can be proved 
(see Ng \& van Dam 2000) that Eq. (2)
leads to:
\begin{equation}
\frac{\sigma_{\omega}}{\omega}
\approx a_o \left(\frac{\omega}{\omega_P}\right)^\alpha,~{\rm or}~
\frac{\sigma_{E}}{E} \approx a_o \left(\frac{E}{E_P}\right)^\alpha.
\end{equation}
The same reasoning also applies to the
intrinsic uncertainty in data on the wavevector
${\bf k}$ (note also that for measurements directly taken by an observer
$\delta \omega$ and $\delta {\bf k}$,
like $\delta t$ and $\delta {\bf r}$,
are uncorrelated errors), for
if any component of ${\bf k}$ could be known to high
accuracy even in the limit of large $p$ we would be able to surpass
the Planck length limitation in distance determination for that direction.
Thus a similar equation may then be formulated as:
\begin{equation}
\frac{\sigma_k}{k} \approx a_o \left(\frac{\omega}{\omega_P}\right)^\alpha
\end{equation}
where $k$ is the magnitude of ${\bf k}$ and the right side is identical to the
previous equation because $\omega = k$ for photons.
Note indeed that
Eqs. (4) and (5) hold good for ultra-relativistic particles as well.

About the value of $\alpha$, the straightforward choice
is $\alpha = 1$, which by Eq. (2) implies $\sigma_t \sim t_P$,
i.e. the most precise clock has uncertainty $\sim t_P$.  Indeed,
$\alpha = 1$ is just the first order term in a power series expansion of
quantum loop gravity.  However, the quantum nature of time at scales
$\leq t_P$ may be
manifested in  other (more contrived) ways.
In particular,  for random walk models of space-time, where
each step has size $t_P$, $\alpha = 1/2$ (Amelino-Camelia 2000).
On the other hand, it was shown
(Ng 2002)
that as a consequence of
the holographic principle
(which states that the maximum degrees of freedom allowed within a
region of space is given by the volume of the region in Planck units, see
Wheeler 1982, Bekenstein 1973, Hawking 1975, 't Hooft 1993, Susskind 1995)
$\sigma_t$ takes the form $\sigma_t/t \sim (t_P/t)^{\frac{2}{3}}$, leading
to $\alpha = 2/3$ in Eqs. (3) and (4).  Such an undertaking also
has the desirable property  (Ng 2002)
that it readily implies a finite lifetime
$\tau$ for
black holes, viz.,  $\tau \sim G^2 m^3/\hbar c^4$, in agreement with
the earlier calculations of Hawking.

Although the choice of $\alpha$
is not unique, the fact that it appears as an exponent means
different values can lead to wildly varying predictions.  Specifically,
even by taking $E = 10^{20}$ eV (i.e. the highest energy particles known,
where Planck scale effects are
still only $\sim (E/E_P)^\alpha \approx 10^{-9 \alpha}$ in
significance),
an increment of $\alpha$ by 0.5  would demand a
detection sensitivity 4.5 orders of
magnitude higher.  The situation gets much worse as $E$ becomes lower.
Thus if an experiment fails to offer confirmation at a
given $\alpha$, one can always raise the value of $\alpha$, and
the search may never end.  Fortunately, however, it turns out that
all of the three scenarios $\alpha = 1/2, 2/3$, and $1$ may be
clinched by the rapid advances in observational astronomy.

\noindent
2. THE PROPAGATION OF LIGHT IN FREE SPACE

How do Eqs. (4) and (5) modify our perception of the radiation dispersion
relation?  By writing the relation as: 
\begin{equation}
\omega^2 - k^2 = 0
\end{equation}
the answer becomes clear - one simply needs to calculate the uncertainty
in $\omega^2 - k^2$ due to the 
intrinsic fluctuations in the measurements of $\omega$ and ${\bf k}$, viz.
$\delta (\omega^2 - k^2) =
2 \omega \delta \omega - 2k \delta k$, 
bearing in mind that $\delta \omega$ and $\delta k$
are independent variations, as already discussed.  This allows us
to obtain the standard deviation
\begin{equation}
\sigma_{\omega^2-k^2} =
2 \sqrt{2} \omega^2 a_o \left(\frac{\omega}{\omega_P}\right)^\alpha
\end{equation}
Thus, typically Eq. (6) will be replaced by:
\begin{eqnarray}
\omega^2 - k^2 \approx \pm 2 \sqrt{2} \omega^2 a_o
\left(\frac{\omega}{\omega_P}\right)^\alpha
\end{eqnarray}
In the case of a positive fluctuation on the right hand term of Eq. (6)
by unit $\sigma$,
the phase and group velocities of propagation will read, for
$E/E_P \ll 1$, as:
\begin{eqnarray}
v_p = \frac{\omega}{k} \approx 
1 + \sqrt{2} a_o \left(\frac{\omega}{\omega_P} \right)^\alpha;~~
v_g = \frac{d \omega}{dk} \approx 1 + \sqrt{2} (1 + \alpha) a_o
\left(\frac{\omega}{\omega_P}\right)^\alpha
\end{eqnarray}
The results differ from that of a particle - here $\omega^2 - k^2$
is a function of $\omega$ and not a constant, so that both
$v_p$ and $v_g$ are $> 1$, i.e. greater than the speed of light $v=1$.
On the other hand, if the right side of Eq. (6) fluctuates negatively
the two wave velocities will read like:
\begin{eqnarray}
v_p = \frac{\omega}{k} \approx 
1 - \sqrt{2} a_o \left(\frac{\omega}{\omega_P} \right)^\alpha;~~
v_g = \frac{d \omega}{dk} \approx 1 - \sqrt{2} (1 + \alpha) a_o
\left(\frac{\omega}{\omega_P} \right)^\alpha
\end{eqnarray}
and will both be $< 1$.

Is it possible to force a re-interpretation of Eq. (8) in another (more
conventional) way, viz., for
a particular off-shell mode $\omega^2 - k^2$ typically assumes a 
{\it constant} value
different from zero by about the unit $\sigma$
of Eq. (7)?  The point, however, is that
even in this (highly artificial)
approach,
the quantities $v_p = \omega/k$ and $v_g = d \omega/dk =
k/\omega$ will still disagree with each other randomly by an amount 
$\sim (\omega/\omega_P)^\alpha$, so that the chief outcome of
Eqs. (9) and (10) is robust.

\vspace{2mm}

\noindent
3. STELLAR INTERFEROMETRY AS AN ACCURATE TEST

But is such an effect observable?  Although an obvious approach is
to employ the highest energy radiation, so as to maximize $\omega/\omega_P$,
such photons are difficult to detect.  More familiar types of
radiation, e.g. optical light,
have much smaller values of $\omega/\omega_P$,
yet the advantage is that we can measure their properties with great accuracy.
Specifically we consider the phase behavior of 1 eV light
received from a
celestial optical source
located at a distance $L$ away.  During the 
propagation time $\Delta t = L/v_g$, the phase has advanced from its
initial value $\phi$ (which we assume to be well-defined)
by an amount:
\begin{eqnarray}
\Delta \phi = 2 \pi \frac{v_p \Delta t}{\lambda} = 2 \pi \frac{v_p}{v_g} 
\frac{L}{\lambda} \nonumber
\end{eqnarray}
According to Eqs. (9) and (10), $\Delta \phi$ should then randomly fluctuate
in the following manner:
\begin{equation}
\Delta \phi =  2 \pi \frac{L}{\lambda} \left[1 \pm \sqrt{2} \alpha a_o
\left(\frac{\omega}{\omega_P}\right)^\alpha \right]
\end{equation}
In the limit when
\begin{equation}
\sqrt{2} \alpha a_o
\left(\frac{\omega}{\omega_P}\right)^\alpha \frac{L}{\lambda} \geq 1;~~
{\rm or}~~ \frac{\sqrt{2} \alpha a_o}{h} E^{1+\alpha} E_P^{-\alpha} L \geq 1
\end{equation}
the phase  of the wave will have appreciable
probability of assuming any value between $0$ and $2 \pi$ upon arrival, 
irrespective of how sharp the initial phase at the source may be.
Since $a_o$ and $\alpha$ are free parameters,
Eqs. (11) and (12) are a common consequence of many quantum
gravity models - both equations can be derived in a variety of ways - though
the approach presented here may be taken as representative.

From the preceding paragraph, a way towards testing the behavior of time
to the limit
has become apparent.  In stellar interferometry
(see e.g. Baldwin \& Haniff 2002 for a review)
light waves from an astronomical source
are incident upon two reflectors (within
a terrestrial telescope) and are subsequently coverged to
form  Young's interference
fringes.  By Eq. (11), however,  we see that if the time ceases to
be exact at the Planck scale
the phase of light from a 
sufficiently distant
source will appear random - when $L$ is large enough to satisfy Eq. (12)
the fringes will disappear.  In fact, 
the value of $L$ at which Eq. (12) holds may readily be calculated
for the case of $\alpha = 2/3$ and $\alpha = 1$,
with the results:
\begin{equation}
L_{\alpha = \frac{2}{3}} \geq 2.47 \times 10^{15} a_o^{-1}
(E/1~{\rm eV})^{-\frac{5}{3}} 
~{\rm cm};~~
L_{\alpha = 1} \geq 7.07 \times 10^{24} a_o^{-1}
(E/1~{\rm eV})^{-2}~{\rm cm}.
\end{equation}
For $a_o =1$ and $E = 1$ eV
these distances correspond respectively to 165 AU (or $8 \times 10^{-4}$ pc)
and 2.3 Mpc.   

\vspace{2mm}

\noindent
4. EXAMINING THE TEST IN MORE DETAIL

Since the subject of our search is no small affair we provide here
an alternative (and closer) view of the proposed experiment.  

In a classical approach to
the `untilted' configuration of
Young's interferometry the phase of a plane wave (from a
distant source) at the position 
${\bf r}$ of the double slit system may be written as
$\omega t - {\bf k \cdot r}$ where $t$ is the arrival time of the wavefront.
The electric fields {\bf ${\cal E}_1$} and 
{\bf ${\cal E}_2$} of the waves at some
point P behind the slits where they subsequently meet may then assume the form:
\begin{equation}
{\bf {\cal E}_1} = {\bf \epsilon_1}  |{\cal E}_1|
e^{i(\omega t - {\bf k \cdot r} + \phi_1)}~~;~~
{\bf {\cal E}_2} = {\bf \epsilon_2} |{\cal E}_2|
e^{i(\omega t - {\bf k \cdot r} + \phi_2)},
\end{equation}
where  $|{\cal E}|$ denotes the modulus of the (complex) magnitude
of vector ${\cal E}$,
${\bf \epsilon_1}$ and ${\bf \epsilon_2}$ are unit vectors, and
$\phi_1$, $\phi_2$ are the advances in the wave phase during the transits
between each of the two slits and P.
The intensity of light at P is proportional
to $|{\bf {\cal E}_1 + {\cal E}_2}|^2$, 
which contains the term essential to the
formation of fringes, viz. $2 |{\cal E}_1| |{\cal E}_2| cos \phi$, where
$\phi = \phi_1 - \phi_2$ and ${\bf \epsilon_1}$, 
${\bf \epsilon_2}$ are taken to be
parallel (as is commonly the case).

If, however, there exist intrinsic and independent
uncertainties in one's knowledge of
the period and wavelength  of light on scales $t_P$ and $c t_P$ respectively,
the most monochromatic plane wave will have to be
a superposition
of many waves, each having slightly varying $\omega$ and ${\bf k}$, and
the phase
velocity $v_p = \omega/k$ will fluctuate according to Eqs. (9) and (10).
For optical frequencies the only measurable effect is the phase separation
between these waves after travelling a large distance at different speeds,
i.e. Eq. (14) will be replaced by:
\begin{eqnarray}
{\bf {\cal E}_1} = {\bf \epsilon_1} |{\cal E}_1| \sum_{j} a_j
e^{i(\omega_j t - {\bf k_j \cdot r} + \phi_1 + \theta_j)} \approx
{\bf \epsilon_1} 
|{\cal E}_1| e^{i(\omega t - {\bf k \cdot r} +\phi_1)} \sum_{j} a_j
e^{i \theta_j}, \nonumber
\end{eqnarray}
and:
\begin{eqnarray}
{\bf {\cal E}_2} = {\bf \epsilon_2} |{\cal E}_2| \sum_{l} a_l
e^{i(\omega_l t - {\bf k_l \cdot r} + \phi_2 + \theta_l)} \approx
{\bf \epsilon_2} 
|{\cal E}_2| e^{i(\omega t - {\bf k \cdot r} +\phi_2)} \sum_{l} a_l
e^{i \theta_l}, \nonumber
\end{eqnarray}
where \{$a_i$\} are real coefficients (not to be confused
with the  $a_o$ of Eq. (1)) normalized such that 
$a_i^2$ equals the occurrence
probability of the 
$i$th wave, which of course is governed by how far $\theta_i$ differs
from its zero mean value when compared with the 
standard deviation in
Eq. (11) (viz. $2 \sqrt{2} \pi \omega/\omega_P$ when $\alpha = 1$).  Note that
this time the intensity at P depends on many `cross' terms, each
of the form $2 a_j a_l |{\cal E}_1| |{\cal E}_2| cos \phi$, where now $\phi =
(\phi_1 - \phi_2) + (\theta_j - \theta_l)$. If the propagation
length $L$ is large enough to satisfy Eq. (12) $\theta_j$ and $\theta_l$,
hence $\theta_j - \theta_l$,
will  spread over one phase cycle,  so that the original
term $2 |{\cal E}_1| |{\cal E}_2| cos (\phi_1 - \phi_2)$  will no longer 
be characteristic of the point P.  This is the mathematical
demonstration of why no appreciable 
fringe contrast across the detector can be expected. Obviously, the
argument can readily be generalized to conclude that if Eq. (12)
is fulfilled interference effects from {\it multiple slits} 
(or a single large slit as limiting case) will
also disappear.

\vspace{2mm}
5. THE DIFFRACTION OF LIGHT FROM EXTRAGALACTIC POINT SOURCES -
ABSENCE OF ANOMALOUS BEHAVIOR IN TIME AND SPACE AT THE PLANCK SCALE

Let us now consider the observations to date.
The Young's type of interference effects were clearly seen
at $\lambda =$ 2.2 $\mu m$ ($E \approx$ 0.56 eV) light from 
a source at 1.012 kpc distance, viz. the star S Ser, 
using the
Infra-red Optical Telescope Array, which enabled a radius determination of
the star (van Belle, Thompson, \& Creech-Eakman 2002).  When comparing
with Eq. (13) we see that this result can already be used to completely
exclude the $\alpha = 2/3$ model, because for such a value of $\alpha$ and
for all reasonable values of $a_o$,
$\Delta \phi$ carries uncertainties $\gg 2 \pi$, and the
light waves would not have interfered.  It is also evident
from Eq. (13), however,
that no statement about $\alpha = 1$ can be made with the
S Ser findings.

Within the context of the previous section's development, it turns out that
the well recognized presence 
of diffraction pattern in the image of extragalactic
point sources when they are viewed through the finite aperture of
a telescope provides even more stringent constraints on $\alpha$.
Such patterns have been observed from sources located at
distances much larger than 1 kpc, implying, as before, that 
phase coherence
of light is maintained at the aperture entrance despite the contrary
prediction of quantum gravity.  In particular,
to clinch the first order prediction
$\alpha=1$ we note that
Airy rings  (circular diffraction) were clearly visible
at both the zeroth and first maxima in an observation
of the active galaxy PKS1413+135 ($L =$ 1.216 Gpc) by
the HST at 1.6 $\mu m$ wavelength (Perlman et al 2002).
Referring back to Eq. (13),
this means exclusion of all $\alpha=1$  quantum gravity fluctuations
that occur at an amplitude $a_o \geq 3.14 \times 10^{-3}$ (moreover the speed
of light does not fluctuate fractionally by more than $\lambda/L$, or several
parts in $10^{32}$).
To facilitate those who wish to explore the implications in full
we offer the following inequality, derivable directly
from Eq. (12), which the reader can use to readily find for
any value of $\alpha$  the range of
$a_o$ still permitted by the PKS1413 result:
\begin{equation}
a_o < 3.14 \times 10^{-3} \frac{(10^{29})^{\alpha - 1}}{\alpha}
\end{equation}
That Eq. (15) is highly sensitive to $\alpha$ has already been
discussed.
Two consequences immediately emerge from Eq. (15): (a)
for any of the $\alpha < 1$ models to survive, they
must involve ridiculously small values of $a_o$, (b) $\alpha > 1$
remains essentially unconstrained.

Further investigation 
of sources that lie beyond PKS1413 will scrutinize the $\alpha \leq 1$
scenarios more tightly than what is already a very stringent
current limit.  More sophisticated ways of pursuing stellar interferometry
are necessary
to test the cases of $\alpha > 1$, or systematic
effects where the dispersion relation of Eq. (6) is modified by
definite rather than randomly varying terms.  

Nevertheless, the
obvious test bed for quantum gravity has indeed been provided by
this Gpc distance source; the outcome is negative.  No doubt one
anticipates interesting propositions on how time and space
may have contrived to leave behind not a trace of their quanta.
Thus, from Michelson-Morley to extragalactic
interferometry 
there remains no direct experimental evidence of any sort that compels us to
abandon the
structureless, etherless space-time advocated by Einstein.
These points, together with in-depth discussions on how
Planck scale phenomenology affects the appearance of the  extragalactic sky,
will be the subject matter of a paper by Ragazzoni and colleagues.

The authors are grateful to Gerard van Belle, Roberto Ragazzoni,
Jonathan Mittaz, Sir Ian Axford, and Lord James
McKenzie of the Hebrides and Outer Isles for discussions.  They also
wish to thank an anonymous referee for helpful comments.

\noindent
{\bf References}

\noindent
Amelino-Camelia, G., 2000, Towards quantum gravity, Proc. of the XXXV
International\\
\indent Winter School on Theor.  Phys., Polanica, Poland, Ed. Jerzy Kowalski-
Glikman.\\
\indent Lecture Notes in Phys., 541, 1, Berlin: Springer-Verlag.\\
\noindent
Amelino-Camelia, G., 2001, Nature, 410, 1065.\\
\noindent
Baldwin, J.E., \& Haniff, C.A., 2002, Phil. Trans. A360, 969.\\
\noindent
Bekenstein, J.D., 1973, Phys. Rev. D., 7, 2333.\\
\noindent
Hawking, S., 1975, Comm. Math. Phys., 43, 199.\\
\noindent
t'Hooft, G., 1993, in {\it Salamfestschrift}, p. 284, Ed. A Ali et al
(World Scientific,\\ 
\indent Singapore).\\
\noindent
Lieu, R., 2002, ApJ, 568, L67.\\
\noindent
Ng, Y. -J., \& van Dam, H., 2000, Found. Phys., 30, 795.\\
\noindent
Ng, Y. -J., Lee, D. -S., Oh, M.C., \& van Dam, H., 2001, Phys. Lett, B507,
236. \\
\noindent
Ng, Y. -J., 2002, Int. J. Mod. Phys., D., in press (to appear in Dec.
special issue), gr-qc/0201022. \\
\noindent
Perlman, E.S., Stocke, J.T., Carilli, C.L., Sugiho, M., Tashiro, M., \\
\indent Madejski, G.,
Wang, Q.D., Conway, J., 2002, AJ, 124, 2401.\\
\noindent
Richichi, A., Bloecker, T., Foy, R., Fraix-Burnet, D., Lopez, B., Malbet, F.,
Stee, P.,\\
\indent von der Luehe, O., \\
Weigelt, G., 2000, Proc. SPIE, 4006, 80.\\
\noindent
Silk, J., 2001, The Big Bang, 3rd ed., W.H. Freeman \& Co.\\
\noindent
Susskind, L., 1995, J. Math. Phys (N.Y.), 36, 6377.\\
\noindent
van Belle, G.T., Thompson, R.R., Creech-Eakman, M.J., 2002, AJ, 124, 1706.\\
\noindent
Wheeler, J., 1982, Int. J. Theor. Phys., 21, 557. \\

\end{document}